*Article*

# Integrated Design and Implementation of Embedded Control Systems with Scilab

**Longhua Ma** [1], **Feng Xia** [2,3,*] **and Zhe Peng** [1]

[1] State Key Laboratory of Industrial Control Technology, Zhejiang University, Hangzhou 310027, P.R. China; E-Mails: lhma@iipc.zju.edu.cn; pengzhe1113@gmail.com

[2] College of Computer Science and Technology, Zhejiang University, Hangzhou 310027, P.R. China

[3] Faculty of Information Technology, Queensland University of Technology, Brisbane QLD 4001, Australia

\* Author to whom correspondence should be addressed; E-mail: f.xia@ieee.org

**Abstract:** Embedded systems are playing an increasingly important role in control engineering. Despite their popularity, embedded systems are generally subject to resource constraints and it is therefore difficult to build complex control systems on embedded platforms. Traditionally, the design and implementation of control systems are often separated, which causes the development of embedded control systems to be highly time-consuming and costly. To address these problems, this paper presents a low-cost, reusable, reconfigurable platform that enables integrated design and implementation of embedded control systems. To minimize the cost, free and open source software packages such as Linux and Scilab are used. Scilab is ported to the embedded ARM-Linux system. The drivers for interfacing Scilab with several communication protocols including serial, Ethernet, and Modbus are developed. Experiments are conducted to test the developed embedded platform. The use of Scilab enables implementation of complex control algorithms on embedded platforms. With the developed platform, it is possible to perform all phases of the development cycle of embedded control systems in a unified environment, thus facilitating the reduction of development time and cost.

# 1. Introduction

With the availability of ever more powerful and cheaper products, the number of embedded devices deployed in the real world has been far greater than that of the various general-purpose computers such as desktop PCs. The evidence includes the fact that of the 9 billion processors manufactured in 2005, less than 2% were used in PCs, Macs, and Unix workstations, while the remainder went into embedded systems [1]. An embedded system is an application-specific computer system that is physically encapsulated by the device it controls. It is generally a part of a larger system and is hidden from end users. There are a few different architectures for embedded processors, such as ARM, PowerPC, x86, MIPS, etc. Some embedded systems have no operating system, while many more run real-time operating systems and complex multithreaded programs. Nowadays embedded systems are used in numerous application areas, for example, aerospace, instrument, industrial control, transportation, military, consumer electronics, and sensor networks. In particular, embedded controllers that implement control functions of various physical processes have become unprecedentedly popular in computer-controlled systems [2-4]. The use of embedded processors has the potential of reducing the size and cost, increasing the reliability, and improving the performance of control systems.

The majority of embedded control systems in use today are implemented on microcontrollers or programmable logic controllers (PLC). Although microcontrollers and programmable logic controllers provide most of the essential features to implement basic control systems, the programming languages for embedded control software have not evolved as in other software technologies [4,5]. A large number of embedded control systems are programmed using special programming languages such as sequential function charts (SFC), function block languages, or ladder diagram languages, which generally provide poor programming structures. On the other hand, the complexity of control software is growing rapidly due to expanding requirements on the system functionalities. As this trend continues, the old way of developing embedded control software is becoming less and less efficient.

There are quite a lot of efforts in both industry and academia to address the above-mentioned problem. One example is the ARTIST2 network of excellence on embedded systems design (http://www.artist-embedded.org). Another example is the CEmACS project (http://www.hamilton.ie/cemacs/) that aims to devise a systematic, modular, model-based approach for designing complex automotive control systems. From a technical point of view, a classical solution for developing complex embedded control software is to use the Matlab/Simulink platform that has been commercially available for many years. For instance, Bucher and Balemi [6] developed a rapid controller prototyping system based on Matlab, Simulink and the Real-Time Workshop toolbox; Chindris and Muresan [7] presented a method for using Simulink along with code generation software to build control applications on programmable system-on-chip devices. However, these solutions are often complicated and expensive. Automatic generation of executable codes directly from Matlab/Simulink models may not always be supported. It is also possible that the generated codes do not perform satisfactorily on embedded platforms, even if the corresponding Matlab/Simulink models are able to achieve very good performance in simulations on PC. Consequently, the developers often have to spend significant time dealing with such situations. As computer hardware is becoming cheaper and cheaper, embedded software dominates the development cost in most cases. In this context, more affordable solutions that use low-cost, even free, software tools rather than expensive proprietary counterparts are preferable.

The main contributions of this paper are multifold. First, a design methodology that features the integration of controller design and its implementation is introduced for embedded control systems. Secondly, a low-cost, reusable, reconfigurable platform is developed for designing and implementing embedded control systems based on Scilab and Linux, which are freely available along with source code. Finally, a case study is conducted to test the performance of the developed platform, with preliminary results presented.

The platform is built on the Cirrus Logic EP9315 (ARM9) development board running a Linux operating system. Since Scilab was originally designed for general-purpose computers such as PCs, we port Scilab to the embedded ARM-Linux platform. To enable data acquisition from sensors and control of physical processes, the drivers for interfacing Scilab with several communication protocols including serial, Ethernet, and Modbus are implemented, respectively. The developed platform has the following main features:

- It enables developers to perform all phases of the development cycle of control systems within a unified environment, thus facilitating rapid development of embedded control software. This has the potential of improving the performance of the resulting system.
- It makes possible to implement complex control strategies on embedded platforms, for example, robust control, model predictive control, optimal control, and online system optimization. With this capability, the embedded platform can be used to control complex physical processes.
- It significantly reduces system development cost thanks to the use of free and open source software packages. Both Scilab and Linux can be freely downloaded from the Internet, thus minimizing the cost of software.

While Scilab has attracted significant attention around the world, limited work has been conducted in applying it to the development/implementation of practically applicable control applications. Bucher *et al.* [8] presented a rapid control prototyping environment based on Scilab/Scicos, where the executable code is automatically generated for Linux RTAI. The generated code runs as a hard real-time user space application on a standard PC. The changes in the Scilab/Scicos environment needed to interface the generated code to the RTAI Linux OS are described. Hladowski *et al.* [9] developed a Scilab-compatible software package for the analysis and control of repetitive processes. The main features of the implemented toolkit include visualization of the process dynamics, system stability analysis, control law design, and a user-friendly interface. Considering a control law designed with Scicos and implemented on a distributed architecture with the SynDEx tool, Ben Gaid *et al.* [10] proposed a design methodology for improving the software development cycle of embedded control systems. Mannori *et al.* [11] presented a complete development chain, from the design tools to the automatic code generation of stand alone embedded control and user interface program, for industrial control systems based on Scilab/Scicos.

The rest of this paper is organized as follows. In the next Section, we introduce the primary software tool used, i.e., Scilab. Section 3 discusses the software design lifecycle in embedded control systems and presents the design methodology adopted in this paper. In Section 4, the implementation of the platform is described. Details of three major components, i.e., hardware, software, and interfaces, are given. The developed system is tested in Section 5 using an illustrative example. Experimental results are presented. We conclude the paper in Section 6.

## 2. The Scilab/Scicos Environment

Scilab (http://www.scilab.org) [12, 13] is a free and open source scientific software package for numerical computations, which provides a powerful open computing environment for engineering and scientific applications. It has been developed by researchers from INRIA and ENPC, France, since 1990 and distributed freely and in open source via the Internet since 1994. It is currently the responsibility of the Scilab Consortium, whch was launched in 2003. Scilab is becoming increasingly popular in both educational/academic and industrial environments worldwide.

Scilab provides hundreds of built-in powerful primitives in the form of mathematical functions. It supports all basic operations on matrices such as addition, multiplication, concatenation, extraction, and transpose, etc. It has an open programming environment in which the user can define new data types and operations on these data types. In particular, it supports a character string type that allows the online creation of functions. It is easy to interface Scilab with Fortran, C, C++, Java, Tck/Tk, LabView, and Maple, for example, to add interactively Fortran or C programs. Scilab has sophisticated and transparent data structures including matrices, lists, polynomials, rational functions, linear systems, among others. It includes a high-level programming language, an interpreter, and a number of toolboxes for linear algebra, signal processing, classic and robust control, optimization, graphs and networks, etc. In addition, a large (and increasing) number of contributions can be downloaded from the Scilab website. The latest stable release of Scilab (version 4.1.2) can work on GNU/Linux, Windows 2000/XP/VISTA, HP-UX, and Mac OS.

**Figure 1.** A screen shot of Scilab/Scicos on a PC.

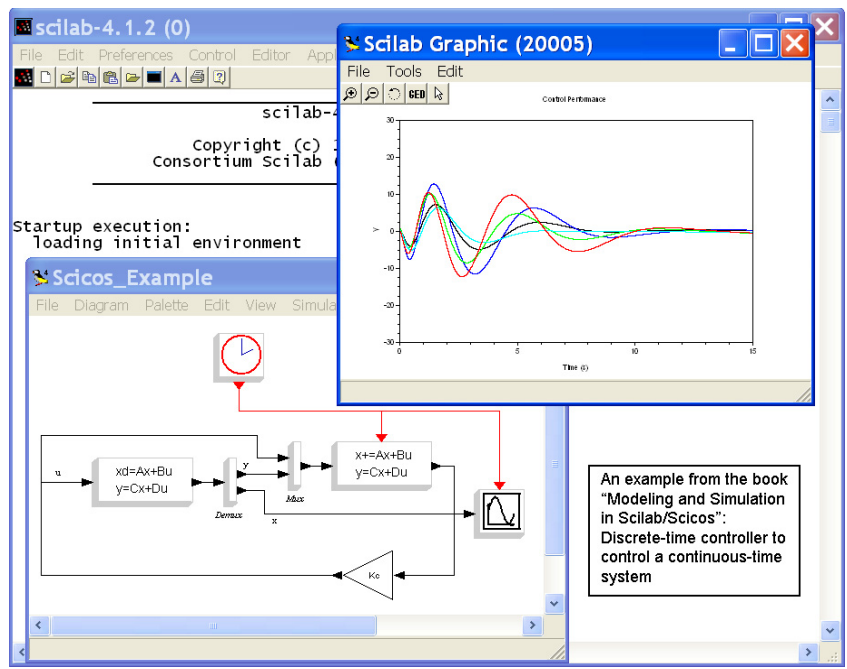

Scilab includes a graphical system modeler and simulator toolbox called Scicos (http://www.scicos.org), which corresponds to Simulink in Matlab. Scicos is particularly useful in signal processing, systems control, and study of queuing, physical, and biological systems. It enables the user to model and simulate the dynamics of hybrid dynamical systems through creating block diagrams using a GUI-based editor and to compile models into executable codes. There are a large number of standard blocks available in the palettes. It is possible for the user to program new blocks in C, Fortran, or Scilab Language and construct a library of reusable blocks that can be used in different systems. Scicos allows running simulations in real time and generating C code from Scicos model using a code generator. Scilab/Scicos is the open source alternative to commercial software packages for system modeling and simulation such as Matlab/Simulink. Figure 1 gives a screen shot of the Scilab/Scicos package.

## 3. Embedded Control Systems Design

### 3.1. Architecture

As control systems increase in complexity and functionality, it becomes impossible in many cases to use analog controllers. At present almost all controllers are digitally implemented on computers. The introduction of computers in the control loop has many advantages [2]. For instance, it makes possible to execute advanced algorithms with complicated computations, and to build user-friendly GUI. The general structure of an embedded control system with one single control loop is shown in Figure 2. The main components consist of the physical process being controlled, a sensor that contains an A/D (Analog-to-Digital) converter, an embedded computer/controller, an actuator that contains a D/A (Digital-to-Analog) converter, and, in some cases, a network.

**Figure 2.** General structure of embedded control systems.

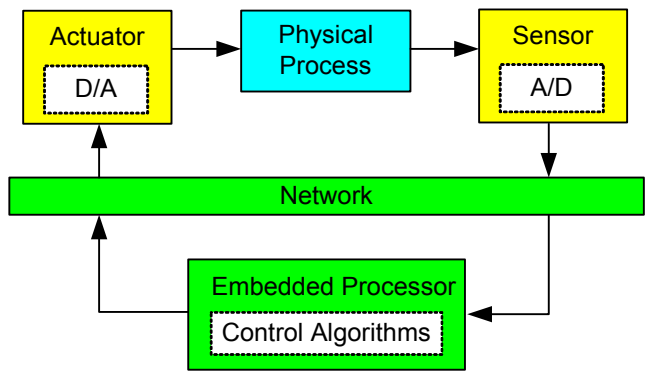

The most basic operations within the control loop are sensing, control, and actuation [3]. The controlled system is usually a continuous-time physical process, e.g. DC motor, inverted pendulum, etc. The inputs and outputs of the process are continuous-time signals. The A/D converter transforms the outputs of the process into digital signals at sampling instants. It can be either a separated unit, or embedded into the sensor. The controller takes charge of executing software programs that process the sequence of sampled data according to specific control algorithms and then produce the sequence of control commands. To make these digital signals applicable to the physical process, the D/A converter transforms them into continuous-time signals with the help of a hold circuit that determines the input to the process until a new control command is available from the controller. The most common method is the zero-order-hold that holds the input constant over the sampling period. In a networked environment, the sequences of sampled data and the control commands need to be transmitted from the sensor to the controller and from the controller to the actuator, respectively, over the communication network. The network could either be wireline (e.g. fieldbus, Ethernet, and Internet) or be wireless (e.g. WLAN, ZigBee, and Bluetooth). In a multitasking/multi-loop environment, as illustrated in Figure 3, different tasks will have to compete for the use of the same embedded processor on which they run concurrently.

**Figure 3.** A multitasking embedded control system.

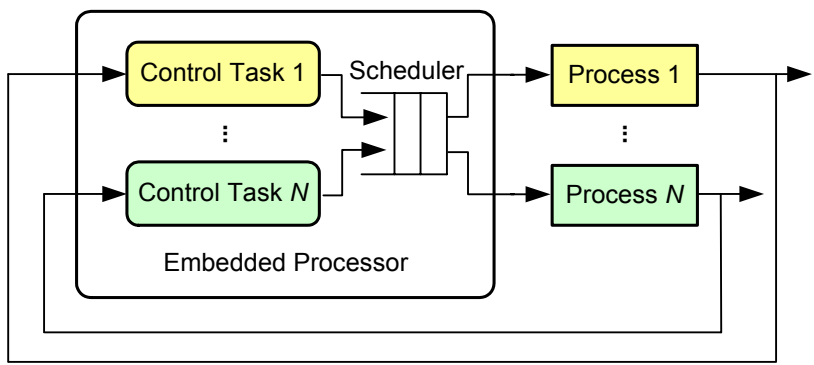

*3.2. Design Methodology*

There is no doubt that embedded control systems constitute an important subclass of real-time systems in which the value of the task depends not only on the correctness of the computation but also on the time at which the results are available [3]. From a real-time systems point of view, the temporal behaviour of a system highly relies on the availability of resources. Therefore, it is compulsory for the system to gain sufficient resources within a certain time interval in order that the execution of individual tasks can be completed in time. Unfortunately, most embedded platforms are suffering from resource limitations, which is in contrast to general-purpose computer systems. There are many reasons behind. For instance, embedded devices are often subject to various limitations on physical factors such as size and weight due to the stringent application requirements. In this context, care must be taken when developing embedded control systems such that the timing requirements of the target application can be satisfied.

Traditionally, the development cycle of a control system consists of two main steps: controller design and its implementation. These two steps are often separated [4, 14], as shown in Figure 4, where the so-called V-model is given. While the controller design is usually done by control engineers, the implementation is the responsibility of system (software) engineers. In the first step, the control engineers model the physical processes using mathematical equations. According to the requirements specification, the control engineers then design the control algorithms. The parameters of the control algorithms are often determined through extensive simulations to achieve the best possible performance. A widely used tool in this step is Matlab/Simulink that supports modeling, synthesis, and simulation of control systems. In this environment the physical processes are usually modeled in continuous time while the control algorithms are discretized to facilitate digital implementation. In the second step, the software engineers produce the programs executing the control algorithms with the parameters designed in the first step. There are a number of mature programming languages available for the implementation. The system will be tested, possibly many times before the satisfactory performance is achieved.

**Figure 4.** Traditional development process of control software.

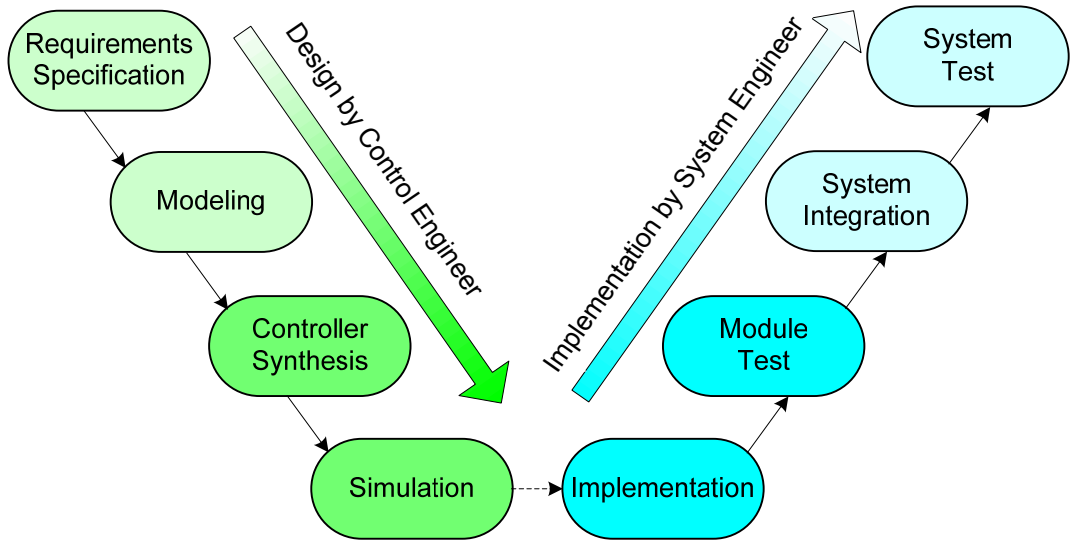

The traditional development process features separation of control and scheduling. The control engineers pay no attention to how the designed control algorithms will be implemented, while the software engineers have no idea about the requirements of the control applications with respect to temporal attributes. In resource-constrained embedded environments, the traditional design methodology cannot guarantee that the desired temporal behavior is achieved, which may lead to much worse-than-possible control performance. Furthermore, the development cycle of a system that can deliver good performance may potentially take a long time, making it difficult to support rapid development that is increasingly important for commercial embedded products.

**Figure 5.** Integrated design and implementation on a unified platform.

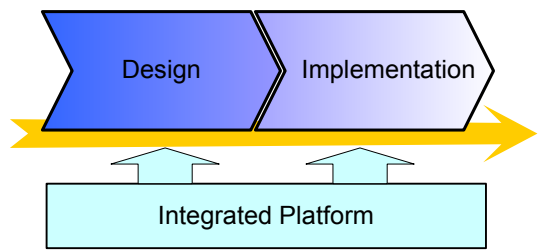

In this paper we adopt a design methodology that bridges the gaps between the traditionally-separated two steps of the development process. As shown in Figure 5, we develop an integrated platform that provides support for all phases of the whole development cycle of embedded control systems. With this platform, the modeling, synthesis, simulation, implementation, and test of control software can be performed in a unified environment. Thanks to the seamless integration of the controller design and its implementation, this design methodology enables rapid development of high-quality embedded controllers that can be used in real-world systems.

## 4. Platform Implementation

In this section, we describe the implementation of the above-mentioned platform for developing embedded control systems. As shown in Figure 6, this platform is composed of three main components: hardware, software, and interfaces. In the following, details of each component are given, respectively.

**Figure 6.** Layered architecture of the developed embedded platform.

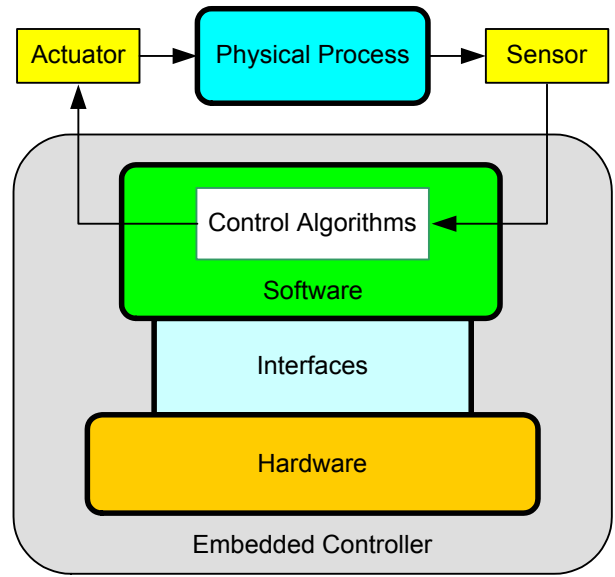

*4.1. Hardware*

The development board used in this work is based on the EP9315 processor from Cirrus Logic, as shown in Figure 7. The EP9315 [15] is a highly integrated system-on-chip processor for consumer and industrial electronic products. It features an advanced 200 MHz ARM920T processor design with a memory management unit, separate 16KB instruction cache, 16KB data cache, 64MB SDRAM, and 32MB flash memory. Linux, Windows CE and many other embedded operating systems are supported. The ARM920T has a 32-bit microcontroller architecture, along with a five-stage pipeline, and is capable of delivering impressive performance at very low power.

The ARM920T core is augmented by the MaverickCrunch coprocessor. This coprocessor greatly accelerates the ARM920T's single- and double-precision integer and floating-point processing capabilities. The board includes a 10/100 Mbps Ethernet media access controller (MAC), a three-port USB 2.0 host, running at 12 Mbps, three UARTs, and external interfaces to SPI, AC97, IIS audio, PCMCIA, Raster/LCD, IDE storage peripherals, keypad and touchscreen, etc. In addition, a LG-Philips LB064V02-TD01 LCD is used to achieve user-friendly display.

**Figure 7.** The development board used.

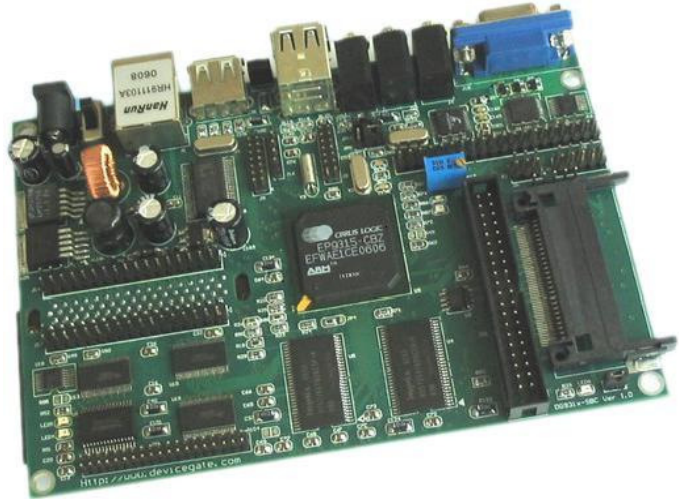

*4.2. Software*

The key software packages used include Linux, TinyX, JWM, and Scilab/Scicos. All these tools can be freely downloaded from the Internet, see Table 1. Linux is a clone of the Unix OS and is most widely used as an operating system in embedded systems. Linux has almost all the features of a modern Unix system. The flexibility, scalability, reliability, and free nature of Linux have made it an increasingly popular platform for a large number of applications. The users can easily remove or modify components of the system that are not needed for a specific embedded system. Linux can run on many different types of processor architectures. While real-time versions of Linux have been available, the standard Linux is adopted in this work primarily because there is no need of writing kernel code. More discussion of the advantages and limitations of using standard Linux for real-time applications can be found in [16].

**Table 1.** Websites for software packages.

| *Software* | *URL* |
|---|---|
| Linux | www.linux.org |
| Scilab | www.scilab.org |
| Scicos | www.scicos.org |
| TinyX | www.xfree86.org |
| JWM | www.joewing.net |

TinyX is a family of X servers designed to be particularly small, which is well suited for embedded systems. TinyX tends to avoid large memory allocations at runtime, and tries to perform operations on-the-fly whenever possible. Unlike the usual XFree86 server, TinyX does not require any configuration files, and will function even if no on-disk fonts are available. With TinyX, the users can easily build their own GUI applications. JWM is a window manager for the X11 window system. It is written in C and uses only Xlib at a minimum.

4.2.1. Embedding Scilab

Scilab/Scicos was originally designed for PC-based systems but not embedded ARM-Linux systems. Therefore, it is necessary to port Scilab/Scicos onto the embedded platform. Since the majority of core codes of Scilab are written in Fortran, we first build a cross-compiler for g77 in order to support cross-compilation of GUI, for example. The GUI system of Scilab/Scicos is based on X11, and therefore the X11 server TinyX is included. To reduce runtime overheads, we optimize/modify some programs in Scilab/Scicos.

We have successfully ported Scilab/Scicos to the ARM-Linux system (see Figure 10). To achieve this goal, a number of files in Scilab and Linux have been modified. The main tasks involved in this process are as follows:

- Port Linux to the ARM platform;
- Port TinyX to ARM-Linux;
- Port JWM to ARM-Linux;
- Port Scilab/Scicos to ARM-Linux;
- Configure and optimize the embedded Scilab/Scicos.

For the first three tasks, technical instructions are available in the literature (see also websites listed in Table 1). Described below are some examples of modifications made for the purpose of porting Scilab/Scicos to ARM-Linux. Detailed programming operations are omitted here for simplicity.

- The `configure` file: on Line 5900, insert
    ```
    int main ()
    {
        return 0;
    }
    ```
- The `configure` file: on Lines 31696 and 31725, delete
    ```
    { (exit 1); exit 1; };
    ```
- `/scilab/routines/xsci/wf_f_util.c`: change Line 76 to
    ```
    extern char *getcwd();
    ```
- `/scilab/routines/xsci/x_misc.c`: insert
    ```
    int sys_nerr;
    char *sys_errlist[];
    ```

4.2.2. Building Control Software in Scilab

Since the source codes of Scilab and Scicos are independent of the underlying system platforms, on the ARM-Linux system it will still be feasible to use programs, blocks, and toolboxes produced on PCs. Scilab has a variety of powerful primitives for programming control applications. Most of them can be found in toolboxes such as *general systems and control* and *robust control toolbox*. There are several different ways to realize a control algorithm in the Scilab/Scicos environment. For instance, it can be programmed as a Scilab `.sci` file written in the Scilab language, or visualized as a Scicos block linked to a specific function/program written in Fortran or C.

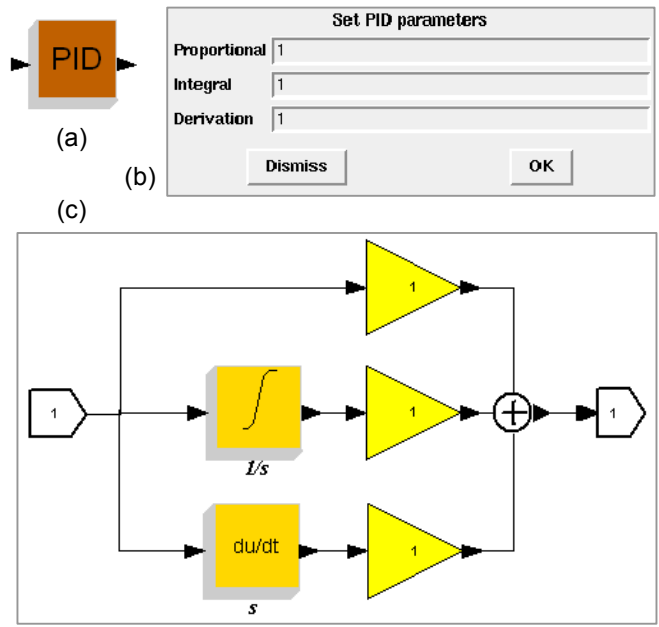

**Figure 8.** A PID controller in Scicos. **(a)** Block. **(b)** Parameter setting dialog box. **(c)** Subsystem.

As a simple example, Figure 8 shows the implementation of a PID controller in Scicos [17]. The PID control algorithm involves three critical parameters, i.e., the Proportional, Integral and Derivative values. The proportional component determines the response to current error. The integral component determines the response with respect to the sum of recent errors. The derivative component determines the response to the change rate of the error. The user can set these three parameters on the dialog box shown in Figure 8(b). The algorithm can be implemented as a (basic) block, see Figure 8(a), or a subsystem that is composed of a number of interconnected blocks, see Figure 8(c).

*4.3. Interfaces*

Scilab supports several ways to interface external code such as Fortran and C programs. One way to call external programs is, for example, to dynamically link the user-developed program with Scilab using the `link` primitive and then to interactively call the linked routine by the `call` primitive. This facilitates the use of specific code, which may be available already from another system or perform better in execution efficiency, in user-defined control software.

In addition to software interfaces, embedded controllers must also provide support for interfacing Scilab with hardware I/O ports. At run time, an embedded controller needs to sample via sensors the output/state of the controlled physical process so as to compute the control command. The corresponding operations will then be performed on the physical process through using actuators. Therefore, Scilab has to communicate with other components in the embedded control system. To address this issue, we developed the drivers, as `.so` files, for interfacing Scilab with serial port, Ethernet, and Modbus on the embedded Linux system. The user interfaces for configuration of the connections are also implemented. These developed I/O interfaces enable not only the basic operations, i.e., sensing and actuation, within control systems, but also the construction of networked, possibly large-scale and complex, control systems. As an illustrative example, Figure 9 describes the procedures for retrieving data from the serial port.

**Figure 9.** Procedures to retrieve data from serial port. (a) Flow chart; (b) Code.

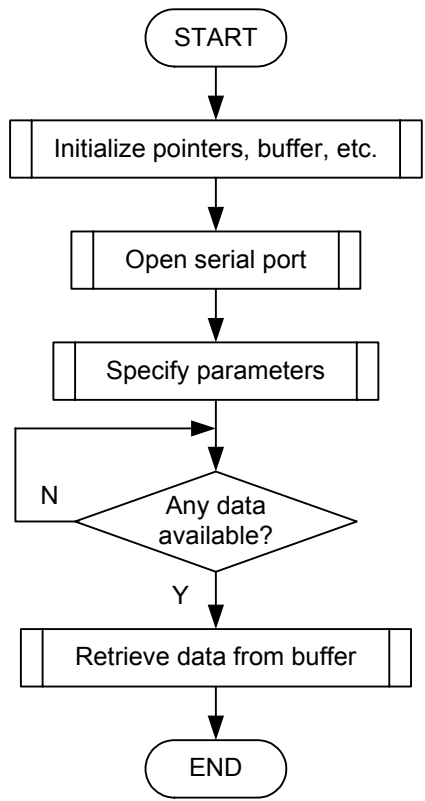

```
int main(int argc, char
**argv){
  int *port=1;
  int *serialspeed=38400;
  int *databits=8;
  int *stopbits=1;
  int *parity=0;
  int *handlevalue;
  int *fd;
  char buff[512];
  *fd =  opencom(port,
  serialspeed, databits,
  stopbits, parity,
  handlevalue);

  while (1)
  {
    Serialread(fd,buff);
    Serialwrite(fd,buff);
  }
  close(fd);
  exit (0);
}
```

(a)                                 (b)

**Figure 10.** The embedded system developed.

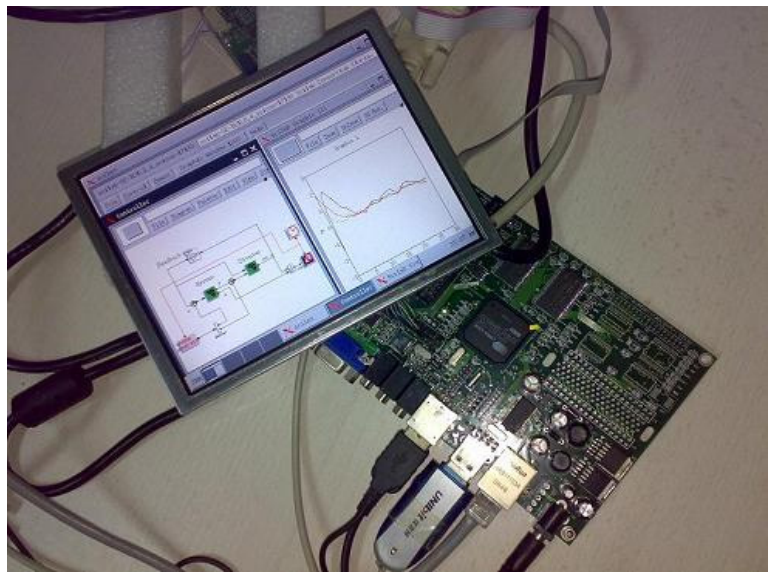

A snapshot of the real system we developed is shown in Figure 10. More details about the implementation of this system can be found in [18].

## 5. Experimental Test

In this section we conduct experiments on the developed embedded control platform. A water tank controller is implemented. The setup of the experimental system is shown in Figure 11. For simplicity the water tank is simulated by a PC running Scilab/Scicos. The PC and the embedded controller are connected using Ethernet. The Scicos models of the water tank running on the PC and the controller running on the ARM-Linux system are depicted in Figures 12 and 13, respectively.

**Figure 11.** Experiment system.

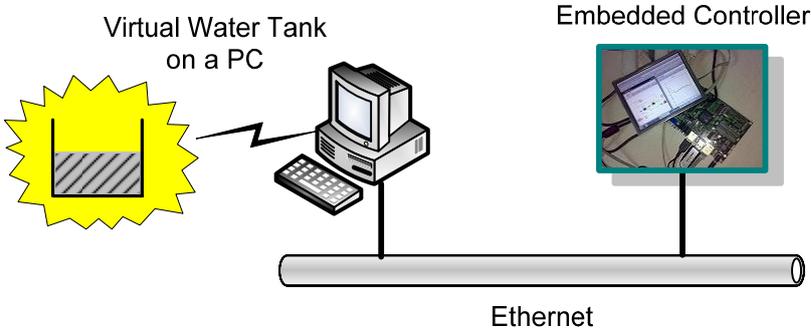

**Figure 12.** Model of water tank.

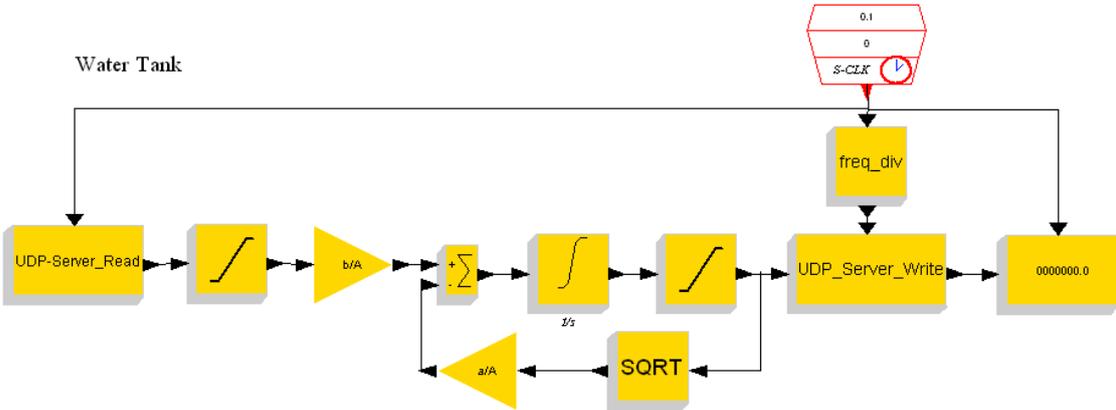

**Figure 13.** Controller on the embedded platform.

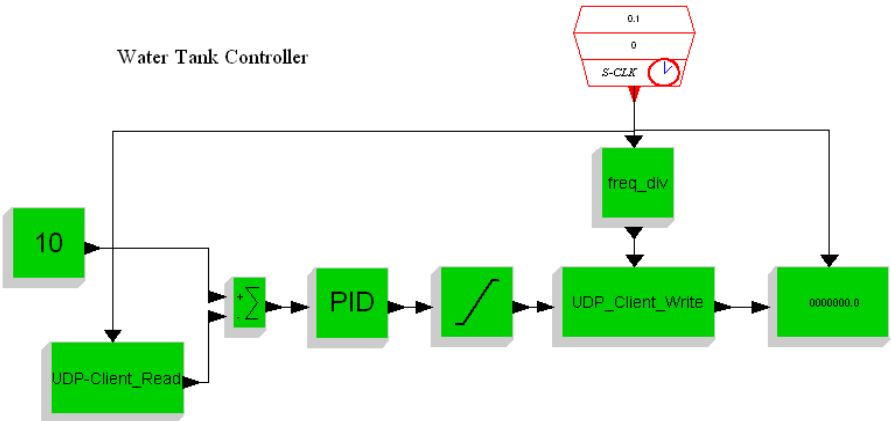

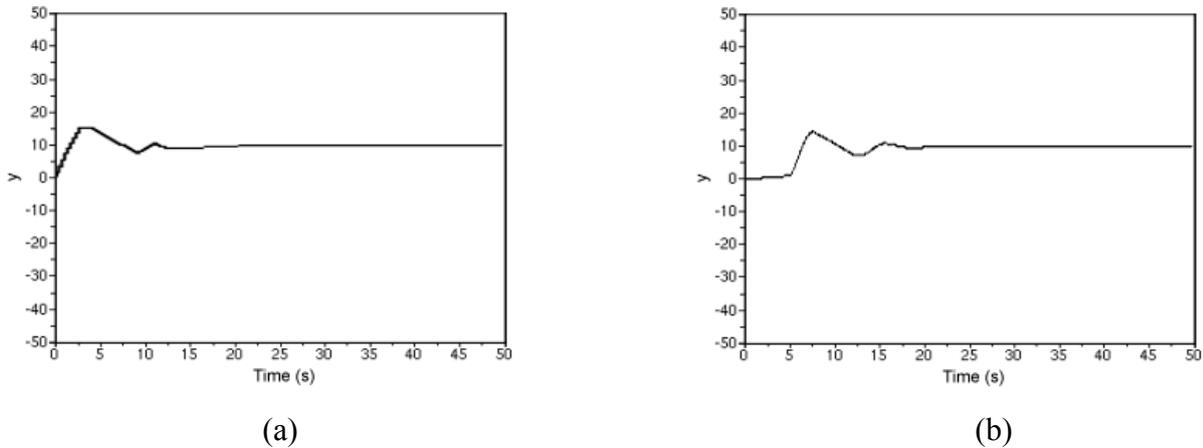

**Figure 14.** Control performance of the experiment system. (a) System output ($h = 0.1$s); (b) System output ($h = 0.5$s).

(a)          (b)

The control performance of the system is shown in Figure 14, where the system output (denoted $y$) is given for different sampling periods, i.e., $h = 0.1$s and $0.5$s, respectively. It is seen that the control system delivers quite good performance, especially when the sampling period is $0.1$s. In both cases, the water level successfully reaches the desired value (i.e. 10 in the experiment) and remains steady after a transient process.

## 6. Conclusions

In this paper we have developed an embedded platform that can be used to design and implement embedded control systems in a rapid and cost-efficient fashion. This platform is built on free and open source software such as Scilab and Linux. Therefore, the system development cost can be minimized. Since the platform provides a unified environment in which the users are able to perform all phases of the development cycle of control systems, the development time can be reduced while the resulting performance may potentially be improved. In addition to industrial control, the platform can also be applied to many other areas such as optimization, image processing, instrument, and education. Our future work includes test and application of the developed platform in real-world systems where real sensors and actuators are deployed.

**Acknowledgements**

This work is supported in part by Natural Science Foundation of China under Grant No. 60474064, Zhejiang Provincial Natural Science Foundation of China under Grant No. Y107476, and China Postdoctoral Science Foundation under Grant No. 20070420232.